\documentclass[aps,prd,groupedaddress,showpacs,eqsecnum]{revtex4}
\usepackage{graphicx}
\usepackage{dcolumn}
\usepackage{bm}
\begin{document}

\title {The effective action and  equations of motion of curved local and global
vortices: Role of the field excitations.}

\author{A.~A.~Kozhevnikov}
\email[]{kozhev@math.nsc.ru} \affiliation{Laboratory of
Theoretical Physics, S.~L.~Sobolev Institute for Mathematics, and
Novosibirsk State University, 630090, Novosibirsk, Russian
Federation}

\date{\today}

\begin{abstract}
The effective actions for both local and global curved vortices
are derived, based on the derivative expansion of the
corresponding field theoretic actions  of the nonrelativistic
Abelian Higgs and Goldstone models. The role of excitations of the
modulus and the phase of the scalar field and of the gauge field
(the Bogolyubov-Anderson mode) emitted and reabsorbed by vortices
is elucidated. In case of the local (gauge) magnetic vortex, they
are necessary for cancellation of the long distance  divergence
when using the transverse form of the electric gauge field
strength of the background field. In case of global vortex taking
them into account results in the Greiter-Wilczek-Witten form of
the effective action for the Goldstone mode. The expressions for
transverse Magnus-like force and the vortex effective mass for
both local and global vortices are found. The equations of motion
of both type of vortices including the terms due to the field
excitations are obtained and solved in cases of large and small
contour displacements.
\end{abstract}
\pacs{67.57.Fg;47.37.+q;11.27.+d;98.80.Cq}

\maketitle

\section{Introduction}
~

The active theoretical studies  of  the superfluid \cite{Pit,Gro}
and superconducting vortices \cite{Abrikos} take place  over more
than half a century \cite{ao03}. Important issues concerning the
dynamics of both types of vortices  include, in particular, the
problems of the transverse Magnus-like force acting on vortex
\cite{thou96,wex96,geller97}, and the evaluation of the vortex
effective mass  \cite{wex96,duan92,Hatsuda96}. The above
incomplete list of references shows that different models are used
in describing various aspects of vortex dynamics. This is due to
complexity of the object under study put in real experimental
conditions which include dissipation, finite-temperature effects
etc.    The  common feature of both types of  vortices in
condensed matter physics is that they appear as topological
defects in the models with spontaneously broken U(1) symmetry.

The purpose of the present paper is to study the dynamics of the
curved line defects  in the framework of a uniform approach based
on the field theoretic models and  to scrutinize  the role of the
excitations of the gauge and scalar fields exchanged between the
different segments of the vortex. The local (gauge) vortex
equations of motion  are obtained from the effective action of the
Abelian Higgs model (AHM)  at zero temperature  $T=0$. The same
task is performed for the global vortex by taking the limit of the
vanishing gauge coupling constant. The earlier attempt to consider
the role of the field excitations was undertaken in
Ref.~\cite{Hatsuda96}. However, the background contribution which
is absolutely essential for the static situation was not taken
into account in \cite{Hatsuda96}. Furthermore,  the treatment in
Ref.~\cite{Hatsuda96}  was restricted to considering only
quasi-two-dimensional situation (straight vortices), and no
attempt was undertaken to derive the vortex equations of motion by
means of the variation of effective action. In the present work we
fill these gaps  and treat the background fields and the
excitations. We work in truly three-dimensional situation and
allow for the curvature of the vortex contour by deriving
expressions for the effective actions which systematically use the
so called derivative expansion. The expressions for the transverse
Magnus-like force and the effective masses of the local and global
vortices are obtained.

\section{Formulation of the model and basic notations.}
\label{basic} ~

We will work in the London limit where the London penetration
depth $\lambda_L$  is much greater than the coherence length
$\xi$. [See Eq.~(\ref{lambdaL}) and (\ref{xi0}) below for the
definition of these quantities.] In this limit the gauge field
configuration of the magnetic vortex  is generated by the singular
phase $\chi_{\rm s}$ of the   order parameter $\psi$
\cite{abrikosov87}. The notion singular means that
\begin{equation}
[{\bm\nabla}\times{\bm\nabla}]\chi_{\rm s}=2\pi\sum_an_a\int
d\sigma_a{\bm X}_a^\prime\delta^{(3)}({\bm x}-{\bm
X}_a),\label{chising}\end{equation} where sum goes over all
contours given by the vector ${\bm X}_a\equiv{\bm
X}_a(t,\sigma_a)$, $n_a$ is the number of flux quanta trapped by
the vortex. Hereafter, prime  means the differentiation with
respect to the corresponding contour parameter $\sigma_a$, and
$[{\bm a}\times{\bm b}]$ stands for the vector product of two
vectors ${\bm a}$ and ${\bm b}$. Since the length of the contour
is $l=\int_{\sigma_{ai}}^{\sigma_{af}}|{\bm
X}^\prime_a|d\sigma_a$, the natural choice of $\sigma_a$ is that
it gives  the length of the contour segment starting with initial
value $\sigma_{ai}$ and ending with $\sigma_a$. Hence, the gauge
condition
\begin{equation}
{\bm X}_a^{\prime2}=1\label{gauge}\end{equation} can be applied.
In the rest of the paper the limits of integration over contour
parameter are omitted for brevity. The action of the Abelian Higgs
model is
\begin{eqnarray}
S&=&\int d^4x\left\{\frac{1}{8\pi}({\bm E}^2-{\bm
H}^2)-\frac{g}{2}(|\psi|^2-n_0)^2+\frac{1}{2}[\psi^\ast(i\hbar\partial_t-q\varphi+qa_0)\psi
+{\rm
c.c.}]-\right.\nonumber\\&&\left.\frac{1}{2m}\left|\left(-i\hbar{\bm\nabla}-\frac{q}{c}{\bm
A}+\frac{q}{c}{\bm a}\right)\psi\right|^2-\rho_0\varphi\right\}.
\label{S}\end{eqnarray}Here,  ${\bm E}=-\partial_t{\bm
A}/c-{\bm\nabla}\varphi$ and  ${\bm H}={\bm\nabla\times{\bm A}}$
are the electric and magnetic field strengths, with $\varphi$
standing for the time component of the gauge four-vector potential
$A_\mu=(\varphi,{\bm A})$; $q=2e$, $m=2m_e$ are the charge, mass,
of the scalar field in terms of the  electronic ones, $c$ is the
velocity of light. Further, $n_0$ and $\rho_0$ stand for the
density of the scalar field condensate (the number density of the
Cooper pairs) and for the homogenous positive charge density
introduced to provide the net neutrality of the system:
\begin{equation}
\rho_0+qn_0=0.\label{neutrality}\end{equation}  The coupling
constant $g$ will be related below to the sound velocity $c_s$.
The four-vector $a_\mu=-\frac{\hbar c}{q}\partial_\mu\chi_s$
represents the four-gradient of the singular phase. The field
strength tensor corresponding to $a_\mu$ is
$f_{\mu\nu}=\partial_\mu a_\nu-\partial_\nu a_\mu\equiv({\bm
E}_{\rm s},{\bm H}_{\rm s})$, where  the mixed Fourier transforms
of the  singular parts of the electric and magnetic field
strengths are
\begin{eqnarray}
{\bm E}_{\rm s}(t,{\bm k})&=&-\frac{\Phi_0}{c}\sum_an_a\int
d\sigma_a[\dot{{\bm X}}_a\times{\bm X}^\prime_a]e^{-i{\bm k}{\bm X}_a},\nonumber\\
{\bm H}_{\rm s}(t,{\bm k})&=&\Phi_0\sum_an_a\int d\sigma_a{\bm
X}^\prime_ae^{-i{\bm k}{\bm X}_a}. \label{singfieldk}
\end{eqnarray}
Hereafter dot means the differentiation with respect to time, and
$$\Phi_0=\frac{2\pi\hbar c}{q}$$is the magnetic flux quantum.
Both the full  Fourier representation in $k=(\omega,{\bm k})$ or
the mixed one in $(t,{\bm k})$ turns out to be  very suitable  in
what follows. The full representation is defined as usual:
\begin{eqnarray}
b_k&\equiv&b(\omega,{\bm k})=\int d^4xe^{i\omega t-i{\bm k}{\bm
x}}b(t,{\bm x}),\nonumber\\
b(t,{\bm x})&=&\int\frac{d^4k}{(2\pi)^4}e^{-i\omega t+i{\bm k}{\bm
x}}b_k.\end{eqnarray} The mixed one is defined accordingly.  One
can find the necessary expressions for the singular part of the
vector potential ${\bm a}_{{\bm k}}(t)\equiv{\bm a}(t,{\bm k})$
from the equation $i[{\bm k}\times{\bm a}_{{\bm k}}]={\bm H}_{\rm
s}(t,{\bm k})$:
\begin{equation}
{\bm a}_{{\bm k}}(t)=i\frac{\Phi_0}{{\bm k}^2}\sum_an_a\int
d\sigma_a[{\bm k}\times{\bm X}^\prime_a]e^{-i{\bm k}{\bm
X}_a}.\label{ak}\end{equation} This vector is transverse, ${\bm
k}{\bm a}_{{\bm k}}=0$. The mixed Fourier component of the
singular part of the Coulomb  potential $a_0$ can be found from
the expression ${\bm E}_{\rm s}(t,{\bm k})=-\dot{{\bm a}}_{{\bm
k}}/c-i{\bm k}a_{0{\bm k}}$:
\begin{equation}
a_{0{\bm k}}(t)=-i\frac{\Phi_0}{c{\bm k}^2}\sum_an_a\int
d\sigma_a{\bm k}[\dot{{\bm X}}_a\times{\bm X}^\prime_a]e^{-i{\bm
k}{\bm X}_a}.\label{a0k}\end{equation} Note that both ${\bm
E}(t,{\bm k})$ and $a_{0{\bm k}}$ look like the local Lorenz
transforms of ${\bm H}(t,{\bm k})$ and ${\bm a}_{{\bm k}}$,
respectively. The  end point contribution vanishes identically for
the closed contour, while for  the open one the end point term is
proportional to the combination $\dot{{\bm
X}}_f\cdot{\bm\nabla}|{\bm x}-{\bm X}_f|^{-1}-\dot{{\bm
X}}_i\cdot{\bm\nabla}|{\bm x}-{\bm X}_i|^{-1}$ which can be made
vanishing either by pinning the end points ${\bm X}_{f,i}$ or
pushing them to the spatial infinity. In what follows we will
always assume the vanishing of the analogous end point
contributions.

\section{The contribution of background field .}
\label{background}~

Let us rewrite the action Eq.~(\ref{S}), first, in terms of the
modulus and the phase of the scalar field $\psi=n^{1/2}e^{i\chi}$.
Second, expand the field configuration in terms of the background
fields and the fluctuations, $\varphi\to \varphi+\delta\varphi$,
${\bm A}\to {\bm A}+\delta{\bm A}$, $n\to n_0+\delta n$,
$\chi\to\delta\chi$. Note that the space-time derivatives of the
singular background phase is already included in Eq.~(\ref{S}) as
the four-vector $a_\mu$. The modulus of the scalar field is taken
to be $n_0$ as it should in the London limit. One gets the total
action Eq.~(\ref{S}) as the sum of the background action $S_{\rm
bg}$ and the action of the fluctuations $S_{\rm f}$. The latter is
discussed in Sec.~\ref{genfunc}. The background action is
\begin{equation}
S_{\rm bg}=\int d^4x\left[\frac{1}{8\pi}({\bm E}^2-{\bm
H}^2)-\frac{1}{8\pi\lambda^2_L}({\bm A}-{\bm
a})^2+n_0qa_0\right].\label{Sbg}\end{equation} The term
proportional to $\varphi$ drops  drops in view of
Eq.~(\ref{neutrality}). Hereafter
\begin{equation}
\lambda_L=\sqrt{\frac{mc^2}{4\pi
n_0q^2}}\label{lambdaL}\end{equation} is the the London
penetration depth. The condition of the vanishing of terms linear
in $\delta{\bm A}$ and $\delta\varphi$ gives the gauge field
(Maxwell) equations for background gauge fields
\begin{equation}
[{\bm\nabla}\times{\bm H}]+\frac{1}{\lambda_L^2}{\bm
A}=\frac{1}{\lambda_L^2}{\bm a}+\frac{1}{c}\partial_t{\bm
E},\label{Maxwell}\end{equation}
\begin{equation}
{\bm\nabla}{\bm E}=4\pi(n_0q+\rho_0)=0,\label{Gauss}\end{equation}
while the vanishing of terms linear in variation of the phase
$\delta\chi$ gives the equation
\begin{equation}
{\bm\nabla}({\bm A}-{\bm
a})=0.\label{transversality}\end{equation} The latter equation is
satisfied automatically in  the Coulomb gauge. Since the situation
is non-relativistic, we will neglect the retardation effects in
what follows. Hence, the displacement current in
Eq.~(\ref{Maxwell}) will be neglected, too. One should not vary
over density because we assume homogeneous density $n=n_0$
everywhere except the vortex core. The latter  is ignored in the
London limit. Using the relation ${\bm H}_{{\bm k}}=i[{\bm
k}\times{\bm A}_{{\bm k}}]$ one can find the solution of
Eq.~(\ref{Maxwell}) in the mixed $(t,{\bm k})$ Fourier
representation:
\begin{equation}
{\bm A}_{{\bm k}}=\frac{1/\lambda^2_L}{{\bm
k}^2+1/\lambda^2_L}{\bm a}_{{\bm
k}}=\frac{i\Phi_0/\lambda_L^2}{{\bm k}^2({\bm
k}^2+1/\lambda^2_L)}\sum_an_a\int d\sigma_a[{\bm k}\times{\bm
X}^\prime_a] e^{-i{\bm k}{\bm X}_a}.\label{Ak}\end{equation}
Making the inverse Fourier transform to the coordinate space one
can see that the vector potential ${\bm A}$ is decomposed into the
long range and the short range terms, while the combination
\begin{equation}
{\bm A}_{{\bm k}}-{\bm a}_{{\bm k}}=-\frac{i\Phi_0}{{\bm
k}^2+1/\lambda^2_L}\sum_an_a\int d\sigma_a[{\bm k}\times{\bm
X}^\prime_a]e^{-i{\bm k}{\bm X}_a}\label{Akmod}\end{equation}
which appears in Eq.~(\ref{Sbg}) contains only the short range
exponentially damped contribution. The expression for the Fourier
amplitude of the strength of the background magnetic field is
\begin{equation}
{\bm H}_{{\bm k}}=\frac{\Phi_0/\lambda^2_L}{{\bm
k}^2+1/\lambda^2_L}\sum_an_a\int d\sigma_a{\bm X}^\prime_a
e^{-i{\bm k}{\bm X}_a}.\label{Hk}
\end{equation}
The Fourier amplitude of the strength of the electric field is
$${\bm E}_{{\bm k}}=-\frac{1/(c\lambda^2_L)}{{\bm
k}^2+1/\lambda^2_L}\dot{{\bm a}}_{{\bm k}}-i{\bm k}\varphi_{\bm
k}.$$ The scalar potential $\varphi_{\bm k}=0$, as can be  found
from the above equation and the Gauss law Eq.~(\ref{Gauss}).  The
resulting expression for ${\bm E}_{{\bm k}}$ looks like
\begin{equation}
{\bm E}_{{\bm k}}=-\frac{\Phi_0/(c\lambda^2_L)}{{\bm
k}^2+1/\lambda^2_L} \sum_an_a\int d\sigma_ae^{-i{\bm k}{\bm
X}_a}\left\{[\dot{{\bm X}}_a\times{\bm X}^\prime_a]-\frac{{\bm
k}({\bm k}[\dot{{\bm X}}_a\times{\bm X}^\prime_a])}{{\bm
k}^2}\right\}.\label{Ek}\end{equation} Both ${\bm H}_{{\bm k}}$
and ${\bm E}_{{\bm k}}$ are transverse.  As for ${\bm E}_{{\bm
k}}$, it is evident from Eq.~(\ref{Ek}). In the case of ${\bm
H}_{{\bm k}}$ the transverse character is proved by noting that
${\bm k}{\bm H}_{{\bm k}}\propto\int d\sigma{\bm k}{\bm X}^\prime
e^{-i{\bm k}{\bm X}}=i\int
d\sigma\frac{\partial}{\partial\sigma}e^{-i{\bm k}{\bm X}}=0$ for
closed contours. For the open ones the transverse character is
provided by pushing the  end points to the  spatial infinity.

With the help of the relation
\begin{equation}
\int d^3xf^2(t,{\bm x})=\int\frac{d^3k}{(2\pi)^3}|f_{{\bm
k}}(t)|^2\label{Parcefal}\end{equation}one can obtain the action
of background fields in terms of the  vortex contour variable
${\bm X}_a\equiv{\bm X}_a(t,\sigma_a)$. From now on we will
restrict ourselves by the case of the single contour with the unit
flux quantum. The notations adopted in what follows are ${\bm
X}_{1,2}\equiv{\bm X}(\sigma_{1,2})$, ${\bm X}_{12}\equiv{\bm
X}_1-{\bm X}_2$, where the points $\sigma_{1,2}$ belong to the
same contour. Taking into account Eqs.~(\ref{Akmod}), (\ref{Hk}),
(\ref{Ek}), and (\ref{Parcefal}) one obtains
\begin{eqnarray}
S_{\rm bg}&=&\frac{\Phi_0^2}{8\pi}\int
dt\frac{d^3k}{(2\pi)^3}\left(\frac{1/\lambda_L^2}{{\bf
k}^2+1/\lambda_L^2}\right)^2\int d\sigma_1d\sigma_2e^{-i{\bm
k}{\bm
X}_{12}}\times\nonumber\\&&\left\{\frac{1}{c^2}\left([{\dot{{\bm
X}}}_1\times{\bm X}^\prime_1][{\dot{{\bm X}}}_2\times{\bm
X}^\prime_2]-\frac{({\bm k}[{\dot{{\bm X}}}_1\times{\bm
X}^\prime_1])({\bm k}[{\dot{{\bm X}}}_2\times{\bm
X}^\prime_2])}{{\bm k}^2}\right)-{\bm X}^\prime_1{\bm
X}^\prime_2)(1+\lambda_L^2{\bm k}^2)\right\}+\nonumber\\&&n_0q\int
d^4xa_0.\label{Sbg1}
\end{eqnarray}In the braces, the terms with the  time derivatives originate from
${\bm E}^2$, the term  $\propto\lambda_L^2$ does from the kinetic
energy of the Higgs field $\propto({\bm A}-{\bm a})^2$, while the
remaining one  is due to ${\bm H}^2$. Notice that the contribution
of the term with $a_0$ in Eq.~(\ref{Sbg1}), at first sight, seems
to be irrelevant  because
$$\int d^3xa_0=-\frac{\Phi_0}{4\pi c}\int d\sigma[{\dot{{\bm
X}}}\times{\bm X}^\prime]\left(\int d^3x{\bm\nabla}\frac{1}{|{\bm
x}-{\bm X}|}\right)=0.$$ However, its variation over ${\bm
X}(\sigma,t)$ is not zero and,  as is found in
Sec.~\ref{localeqmo}, results in the Magnus-like transverse force.

Let us obtain the  leading and next-to-leading terms in the
expansion over  small parameter $\kappa^2\lambda^2_L$, where
$\kappa=|{\bm X}^{\prime\prime}|$ is the curvature of the contour.
The dominant contributions come from the nearby points of the
contour $\sigma_2=\sigma_1+z$, $z\ll\sigma_1$, hence the
derivative expansion
\begin{eqnarray}
{\bm X}_2&=&{\bm X}_1+z{\bm X}^\prime_1+\frac{z^2}{2}{\bm
X}^{\prime\prime}_1+\frac{z^3}{6}{\bm X}^{\prime\prime\prime}_1+\cdots,\nonumber\\
{\bm X}_{21}&=&z{\bm X}^\prime_1+\frac{z^2}{2}{\bm
X}^{\prime\prime}_1+\frac{z^3}{6}{\bm X}^{\prime\prime\prime}_1+\cdots,\nonumber\\
{\bm X}^\prime_2&=&{\bm X}^\prime_1+z{\bm
X}^{\prime\prime}_1+\frac{z^2}{2}{\bm
X}^{\prime\prime\prime}_1+\cdots,\nonumber\\
|{\bm X}_{21}|&=&|z|\left(1-\frac{z^2}{24}\kappa^2+\cdots\right)
\label{expansion}\end{eqnarray} can be used. When obtaining the
bottom  line in Eq.~(\ref{expansion}) the relations ${\bm
X}^\prime\cdot{\bm X}^{\prime\prime\prime}=-{\bm
X}^{\prime\prime2}=-\kappa^2$ and ${\bm X}^\prime\cdot{\bm
X}^{\prime\prime}=0$ should be taken into account.  They follow
from the Frenet-Serre equations
\begin{eqnarray}
{\bm X}^{\prime\prime}&=&\kappa{\bm n},\nonumber\\
{\bm n}^\prime&=&-\kappa{\bm X}^\prime+\tau{\bm b},\nonumber\\
{\bm b}^\prime&=&-\tau{\bm n}, \label{frenet}
\end{eqnarray}
where $\tau$ stands for the torsion of the contour, and ${\bm n}$,
${\bm b}$ are the vectors of normal and bi-normal, respectively.
The  vectors $({\bm n},{\bm b},{\bm X}^\prime)$ comprise the right
triple of the unit orthogonal vectors, so that ${\bm
X}^\prime=[{\bm n}\times{\bm b}]$ (and  similar relations obtained
by the cyclic permutation). Note also that ${\bm
X}_2^{\prime2}=1+O(z^4)$ if ${\bm X}_1^{\prime2}=1$, hence the
approximate expansions over $z$ do not break the gauge condition
${\bm X}^{\prime2}=1$ within the adopted accuracy.  The background
action is represented as $S_{\rm bg}=S_{\rm bg}^{(0)}+\Delta
S_{\rm bg},$ where
\begin{eqnarray}
S_{\rm bg}^{(0)}&=&\frac{\Phi_0^2}{8\pi}\int
dt\frac{d^3k}{(2\pi)^3}\left(\frac{1/\lambda_L^2}{{\bf
k}^2+1/\lambda_L^2}\right)^2\int d\sigma_1d\sigma_2e^{-i{\bm
k}{\bm X}_{12}}\left\{[{\dot{{\bm X}}}_1\times{\bm
X}^\prime_1]\cdot[{\dot{{\bm X}}}_2\times{\bm
X}^\prime_2]/c^2-({\bm X}^\prime_1{\bm
X}^\prime_2)(1+\lambda_L^2{\bm k}^2)\right\}+\nonumber\\&&n_0q\int
d^4xa_0.\label{S0bg}\end{eqnarray}It is  finite at large distances
while
\begin{eqnarray}
\Delta S_{\rm bg}&=&-\frac{\Phi_0^2}{8\pi c^2}\int
dt\int\frac{d^3k}{(2\pi)^3{\bm
k}^2}\left(\frac{1/\lambda_L^2}{{\bf
k}^2+1/\lambda_L^2}\right)^2\int d\sigma_1d\sigma_2e^{-i{\bm
k}{\bm X}_{12}}({\bm k}[{\dot{{\bm X}}}_1\times{\bm
X}^\prime_1])({\bm k}[{\dot{{\bm X}}}_2\times{\bm
X}^\prime_2])\label{DSbg}\end{eqnarray} diverges at large
distances in view of  the factor $1/{\bm k}^2$. As it will be
shown below, the similar divergent  contribution   with the
opposite sign arises from the term originating form the integrated
out fluctuations. Hence, we postpone the evaluation of the terms
$\propto 1/{\bm k}^2$ to Sec.~\ref{effeact}.

Performing  the integration  over momentum ${\bm k}$ in
Eq.~(\ref{S0bg}) one gets
\begin{eqnarray}
S_{\rm bg}^{(0)}&=&\frac{\Phi^2_0}{32\pi^2\lambda^2_L}\int
dtd\sigma_1d\sigma_2e^{-|{\bm
X}_{21}|/\lambda_L}\left\{\frac{[\dot{{\bm X}_1}\times{\bm
X}_1^\prime][\dot{{\bm X}_2}\times{\bm
X}_2^\prime]}{2c^2\lambda_L}-\frac{({\bm X}^\prime_1{\bm
X}^\prime_2)}{|{\bm X}_{21}|}\right\}+n_0q\int
d^4xa_0.\label{S0bg1}\end{eqnarray}The integrand in
Eq.~(\ref{S0bg1}) is nonlocal in the contour parameter $\sigma$.
This is due to the distribution of the  gauge vortex profile over
coordinate space. However, the  contributions of the remote points
are suppressed exponentially as $e^{-|{\bm X}(\sigma_2)-{\bm
X}(\sigma_2)|/\lambda_L}$, and one can derive the approximate
local expression. Indeed, using $\int d\sigma_1\int
d\sigma_2\approx\int d\sigma_1\int_{-\infty}^\infty dz$ and
replacing $\sigma_1\to\sigma$ one can integrate over $z$ to obtain
\begin{eqnarray}
S_{\rm bg}^{(0)}&=&\left(\frac{\Phi_0}{4\pi\lambda_L}\right)^2\int
dt d\sigma\left\{\frac{1}{2c^2}\left([\dot{{\bm X}}\times{\bm
X}^\prime]^2-\lambda^2_L[\dot{{\bm X}}\times{\bm
X}^\prime]^{\prime 2}\right)-{\bm X}^{\prime
2}\left(\ln\frac{\lambda_L}{\xi}-C-\frac{13}{24}\lambda^2_L\kappa^2\right)\right\}
+\nonumber\\&&n_0q\int d^4xa_0.\label{Sbg0}
\end{eqnarray}where
$C=0.577215....$ is the Euler constant. Notice that the term with
$\propto\ln\lambda_l/\xi$ arises due to the usual short distance
cutoff when the lower integration limit over $z$ is replaced by
the coherence length $\xi$:
\begin{eqnarray}
\int_{-\infty}^\infty dz\frac{e^{-|z|/\lambda_L}}{|z|}
&\to&2\int_\xi^\infty\frac{dz}{z}e^{-z/\lambda_L}
\approx2\left(\ln\frac{\lambda_L}{\xi}-C\right).\label{ultraviol}\end{eqnarray}
Such a replacement is justifiable because the modulus of the
scalar field $|\psi|$, in fact,  vanishes at the vortex core
$r_\bot\leq\xi$. Thus,  the only contribution to the action
enhanced as $\ln\lambda_L/\xi$ is that of the kinetic energy of
the scalar field given by the term $\propto{\bm k}^2$ in
Eq.~(\ref{Sbg1}). It  is recognized as the contribution of the
minus energy of the gauge vortex with the energy per unit length
$\epsilon_{\rm v}$ given, in the leading logarithmic approximation
and by neglecting higher derivative term $\propto\kappa^2$, by the
expression \cite{abrikosov87}
\begin{equation}
\epsilon_{\rm
v}\approx\left(\frac{\Phi_0}{4\pi\lambda_L}\right)^2\ln\frac{\lambda_L}{\xi}.\label{ev}\end{equation}
One should be cautious when dealing with the  approximate local
form Eq.~(\ref{Sbg0}) [together with  the correction given by
Eq.~(\ref{DSeff3})]. Such local expressions are useful only when
identifying various contributions, as, for example, the energy per
unit length above, or  in deriving the expression for the
effective mass of the vortex (see Sec.~\ref{localeqmo},
\ref{global}, and Ref.~\cite{Hatsuda96}). They {\it cannot} be
used for obtaining equations of the vortex motion because of the
above mentioned finite size vortex gauge field distribution
$\propto e^{-|{\bm X}(\sigma_2)-{\bm X}(\sigma_2)|/\lambda_L}$.
Indeed, the variation of the approximate local expression
$\propto\int dtd\sigma{\bm X}^{\prime2}$ in Eq.~(\ref{Sbg0}) gives
the result which is by the factor two greater than the correct
expression obtained upon varying the corresponding nonlocal
contribution $\propto\int d\sigma_1d\sigma_2{\bm X}^\prime_1{\bm
X}^\prime_2e^{-|{\bm X}_{21}|/\lambda_L}$ in Eq.~(\ref{S0bg1}).
The same is true for the dynamical terms containing the velocity
$\dot{{\bm X}}$. See Sec.~\ref{localeqmo} for more detail. To
summarize, only expressions Eq.~(\ref{S0bg1}) [and (\ref{DSeff2})
in Sec.~\ref{localeqmo}] should be used to obtain the vortex
equations of motion.

\section{Integrating out the excitations of the  scalar and gauge  fields. }
\label{genfunc}~

The next step is to integrate out the propagating fluctuations of
the modulus of the order parameter $\delta n$, the phase of this
parameter $\delta\chi$, the Coulomb  potential $\delta\varphi$,
and the vector potential $\delta{\bm A}$. This is necessary for
obtaining the correction to the action of the background fields.
To this end let us write the action $S_{\rm f}$ for fluctuations
representing them for short as the row $f^T=(\delta
n,\delta\chi,\delta\varphi)$  and $\delta{\bm A}$.  One obtains
$S_{\rm f}=S_{\rm f}^{(0)}+S_{\rm f}^{(1)}$, where
\begin{eqnarray}
S_{\rm f}^{(0)}&=&\int
d^4x\left\{\frac{1}{2}f^TMf-\frac{1}{8\pi}\delta{\bm
A}\left(\frac{1}{c^2}\partial^2_t+{\bm\nabla}^2-\frac{1}{\lambda^2_L}\right)
\delta{\bm A}+\delta n\left[qa_0-\frac{q^2}{2mc^2}({\bm A}-{\bm
a})^2\right]\right\}\label{Sf}\end{eqnarray}and
\begin{equation}
S_{\rm f}^{(1)}=\frac{q\hbar}{mc}\int d^4x({\bm A}-{\bm a})\delta
n\left({\bm\nabla}\delta\chi-\frac{q}{\hbar c}\delta{\bm
A}\right).\label{DeltaS}\end{equation} The Coulomb gauge ${\bm
\nabla}\delta{\bm A}=0$ is chosen. The matrix $M$ is
\begin{equation}\left(%
\begin{array}{ccc}
  \frac{\hbar^2}{4mn_0}{\bm\nabla}^2-g & -\hbar\partial_t & -q \\
  \hbar\partial_t & \frac{\hbar^2n_0}{m}{\bm\nabla}^2 & 0 \\
  -q & 0 & -\frac{1}{4\pi}{\bm\nabla}^2 \\
\end{array}%
\right).\label{M}\end{equation}  Formally, $S_{\rm f}^{(1)}$ in
Eq.~(\ref{DeltaS}) is quadratic in fluctuations and should be
attributed to the free action. However, because of  highly
inhomogeneous profile of the background vector potential ${\bm
A}-{\bm a}$, the path integrations with $S_{\rm f}$ cannot be
performed in the closed form, so we prefer to treat $S_{\rm
f}^{(1)}$ as the perturbation leaving $S_{\rm f}^{(0)}$ in
Eq.~(\ref{Sf}) as the action of free fluctuations. The
contribution of the term $q^2({\bm A}-{\bm a})^2/2mc^2$ can be
neglected. In fact, it would enter the effective action either
quadratically giving the terms $\propto\int\int\int\int
d\sigma_1d\sigma_2d\sigma_3d\sigma_4X^\prime_1X^\prime_2X^\prime_3X^\prime_4$,
or as the interference with $qa_0$ giving the terms
$\propto\int\int\int
d\sigma_1d\sigma_2d\sigma_3\dot{X}_1X^\prime_1X^\prime_2X^\prime_3$.
Both type of the terms  are attributed to  higher derivatives and
go beyond the scope of the present paper. One can show also that
keeping  $S_{\rm f}^{(1)}$ results in the higher derivative terms
in the effective action. See the end of the present section.

As is known,  one should find the generating functional $Z[j]$ in
order to integrate out the quantum excitations \cite{ramond82}. In
the present case it looks as
\begin{equation}
Z[j]=\int D[\delta n]D[\delta\chi]D[\delta \varphi]D[\delta{\bm
A}]\exp[\frac{i}{\hbar}\widetilde{S_{\rm f}}]
\label{Z}\end{equation} where the action $\widetilde{S_{\rm
f}}=S_{\rm f}^{(0)}+\int d^4x(j_f^Tf+{\bf j}_{{\bm A}}\delta{\bm
A})$ is the functional of currents $j^T=(j_n,j_\chi,j_{\varphi})$,
and ${\bf j}_{{\bm A}}$. The gauge fixing and ghost terms are
omitted  because they are irrelevant in the context of the paper.
As in the preceding section, it is suitable to work in the
momentum space $k=(\omega,{\bm k})$. Then the integration measure
for the spacetime function $f(x)$ is expressed in terms of its
Fourier amplitudes as
$$D[f]=\prod_k\frac{d{\rm Re}f_kd{\rm Im}f_k}{2\pi}.$$
The path integral Eq.~(\ref{Z}) is gaussian, so using the inverse
of $M$ in the momentum space,
\begin{equation}
M^{-1}=\frac{1}{\left(\frac{\epsilon^2_B}
{\hbar^2}+\frac{c^2}{\lambda^2_L}-\omega^2-i0\right)}\left(%
\begin{array}{ccc}
  -\frac{n_0}{m}{\bm k}^2 & -\frac{i\omega}{\hbar} & -\frac{4\pi n_0 q}{m} \\
   \frac{i\omega}{\hbar}& -\frac{m}{\hbar^2{\bm k}^2n_0}
   \left(\frac{\epsilon^2_B}{\hbar^2}+\frac{c^2}{\lambda^2_L}\right) & \frac{4\pi i\omega q}{\hbar{\bm k}^2} \\
  -\frac{4\pi n_0 q}{m} & -\frac{4\pi i\omega q}{\hbar{\bm k}^2} & \frac{4\pi}{{\bm k}^2}
  \left(\frac{\epsilon^2_B}{\hbar^2}-\omega^2\right) \\
\end{array}%
\right), \label{Minverse}\end{equation} one finds the generating
functional $Z[j]=\exp\frac{i}{\hbar}\Delta\widetilde{S}_{\rm
f},\label{Zj}$ where  the correction due to excitations is
\begin{eqnarray}
\Delta\widetilde{S}_{\rm f}&=&i\hbar\ln\det M
+\frac{1}{2}\int\frac{d^4k}{(2\pi)^4}\left\{\left(\frac{\epsilon^2_B}
{\hbar^2}+\frac{c^2}{\lambda^2_L}-\omega^2-i0\right)^{-1}\times\right.\nonumber\\&&\left.\left[\frac{n_0{\bm
k}^2}{m}|(j_n)_k+qa_{0k}|^2+\frac{m|(j_\chi)_k|^2}{\hbar^2{\bm
k}^2n_0}\left(\frac{\epsilon^2_B}{\hbar^2}+\frac{c^2}{\lambda^2_L}\right)
-\right.\right.\nonumber\\&&\left.\left.\frac{4\pi|(j_{\varphi})_k|^2}{{\bm
k}^2}\left(\frac{\epsilon^2_B}{\hbar^2}-\omega^2\right)+\frac{i\omega}{\hbar}
[(j_n+qa_0)_k^\ast
(j_\chi)_k-c.c.]+\right.\right.\nonumber\\&&\left.\left.\frac{4\pi
n_0q}{m}[(j_n+qa_0)_k^\ast (j_{\varphi})_k+c.c.]+\frac{4\pi
i\omega}{\hbar{\bm
k}^2}[(j_\chi)^\ast_k(j_{\varphi})_k+c.c.]\right]+\right.\nonumber\\&&\left.\frac{4\pi|({\bm
j}_{{\bm A}})_k|^2}{{\bm
k}^2+\frac{1}{\lambda^2_L}-\frac{\omega^2}{c^2}-i0}\right\}.
\label{Seffk}
\end{eqnarray}
Here,
\begin{equation}
\epsilon^2_B\equiv\epsilon^2_B({\bm k})=\left(\frac{\hbar^2{\bm
k}^2}{2m}\right)^2+\hbar^2c^2_s{\bm
k}^2\label{Bogol}\end{equation} is the square of the Bogolyubov
spectrum \cite{Bogolyubov} looking at  small wave numbers as the
spectrum of sound waves with the sound velocity
\begin{equation}
c_s=\sqrt{\frac{n_0g}{m}}.\label{cs}\end{equation} As is known,
the regularization prescription $-i0$ guarantees that the causal
(Feynman) Green functions will result after taking the variational
derivatives of $Z[j]$ over $j$. The correlation functions in the
momentum space which are just the Fourier transforms of the Green
functions are
\begin{equation}
\langle
f^{(1)}_{k_1}f^{(2)\ast}_{k_2}\cdots\rangle=Z^{-1}[0]\left(-2i\hbar\frac{\delta}
{\delta j^{(1)\ast}_{k_1}}\right)\left(-2i\hbar\frac{\delta}
{\delta j^{(2)}_{k_2}}\right)\cdots
Z[j]|_{j=0}.\label{correlator}\end{equation} The divergent
constant $\ln\det M$ drops from all expressions for the
correlators Eq.~(\ref{correlator}).

The meaning of the poles in $\omega$ in Eq.~(\ref{Minverse})  is
the following. Let us  consider  the freely propagating
fluctuations decoupled from their source represented by the term
in square brackets of Eq.~(\ref{Sf}). Their equations of motion
obtained upon the condition of the vanishing  variational
derivatives look as
\begin{eqnarray}
0&=&\hbar\partial_t\delta\chi+g\delta
n+q\delta\varphi-\frac{\hbar^2}{4mn_0}{\bm\nabla}^2\delta
n,\nonumber\\0&=&\partial_t\delta n+\frac{\hbar n_0}{m}{\bm\nabla}^2\delta\chi,\nonumber\\
0&=&{\bm\nabla}^2\delta\varphi+4\pi q\delta
n.\label{Flucteqmot}\end{eqnarray} Applying ${\bm\nabla}^2$ to the
first line in Eq.~(\ref{Flucteqmot}) and using the second and
third lines of the same equation one finds
$$\left(-\partial^2_t+c^2_s{\bm\nabla}^2-\frac{c^2}{\lambda^2_L}-\frac{\hbar^2}{4m^2}{\bm\nabla}^4\right)\delta
n=0.$$ The propagating plane wave solution has the dispersion law
with the gap
$$\omega^2\equiv\omega^2_{\bm k}=\frac{c^2}{\lambda^2_L}+\frac{\varepsilon^2_B({\bm k})}{\hbar^2},$$
which is just the pole position of the matrix $M^{-1}$. The same
dispersion law is obtained for the phase fluctuations
$\delta\chi$.  The above dispersion law in the long wave limit
acquires the form $\omega^2\approx\omega^2_p+c^2_s{\bm
k}^2=\omega^2_p+\frac{1}{3}v^2_F{\bm k}^2$, where $\omega_p$,
$v_F$ stand for the plasma frequency and Fermi velocity,
respectively. This is the dispersion law for the
Bogolyubov-Anderson mode in the charged Fermi system
\cite{Bog59,anderson}. The sound velocity $c_s$ is related to
$v_F$ through the expression for the sound velocity
$c^2_s=\frac{N}{m}\left(\frac{\partial\mu}{\partial N}\right)_S$
\cite{LifPit},  by taking into account the expression for chemical
potential of the weakly interacting Bose gas $\mu=gn_0/m$
\cite{Bogolyubov,LifPit}. Recall that AHM mimics the Bose liquid
of the Cooper pairs whose parameters are related to its electronic
counterparts by the relations $\mu=2\mu_e$, $n_0=n_e/2$, $m=2m_e$,
$q=2e$. In the case of neutral superfluid where the gauge coupling
constant is switched off, $q\to0$, the spectrum is gapless,
$\omega^2=\varepsilon_B^2/\hbar^2$ \cite{Bogolyubov}.  On the
other hand, taking the static limit and neglecting the gauge
fields one finds the equation
$${\bm\nabla}^2\delta n-\frac{4mn_0g}{\hbar^2}\delta n=0,$$
which means that the space fluctuations of the modulus of the
scalar field damp at the  coherence (or healing) length
\begin{equation}
\xi=\frac{\hbar}{2(mn_0g)^{1/2}}.\label{xi0}\end{equation} Taking
into account Eq.~(\ref{cs}) one obtains the relation
\begin{equation}
\xi=\frac{\hbar}{2mc_s}\label{xi}\end{equation} which relates the
sound velocity $c_s$ with the coherence length $\xi$. We dwell
upon in this paragraph  on the well known issues in order to show
that the character of excitations and their dispersion laws can be
established in the framework of effective lagrangians without
references to the underlying microscopic picture.

The contribution of the term Eq.~(\ref{DeltaS}) can be evaluated
with the help of  Eq.~(\ref{Seffk}). One finds the correction to
the effective action due to the above term:
\begin{eqnarray*}
\Delta S^{(1)}_{\rm
eff}&=&-\frac{q^3n_0}{4m^2\hbar^2}\int\frac{d^4k_1d^4k_2}{(2\pi)^8}({\bm
A}-{\bm a})^\ast_{k_1+k_2}\omega_1{\bm k}_1{\bm
k}^2_2a_{0k_1}a_{0k_2}\left(\frac{\epsilon^2_B({\bm k}_1)}
{\hbar^2}+\frac{c^2}{\lambda^2_L}-\omega^2_1-i0\right)^{-1}\times\nonumber\\&&\left(\frac{\epsilon^2_B({\bm
k}_2)}
{\hbar^2}+\frac{c^2}{\lambda^2_L}-\omega^2_2-i0\right)^{-1}.
\end{eqnarray*}Using Eq.~(\ref{a0k}) and  (\ref{Akmod}) one can
see that because of the factor $\omega_1$ in the integrand the
above expression starts with the terms of the type $\int
d\sigma_1d\sigma_2d\sigma_3X^\prime_1X^\prime_2\ddot{X_3}X^\prime_3$
etc. resulting in the third order  time derivative terms in the
vortex  equations of motion. Such higher derivative terms can be
neglected for sufficiently  smooth evolution of the contours. By
this reason the term Eq.~(\ref{DeltaS}) can be neglected.

\section{Derivative expansion and the effective action for the
local vortex.}\label{effeact}~

The correction due to fluctuations is obtained from
Eq.~(\ref{Seffk}) by setting $j_n=0,j_\chi=0,j_\phi=0$ and
neglecting the term $\ln\det M$:
\begin{equation}
\Delta S_{\rm
f}=\frac{1}{2}\int\frac{d^4k}{(2\pi)^4}\left(\frac{\epsilon^2_B}
{\hbar^2}+\frac{c^2}{\lambda^2_L}-\omega^2-i0\right)^{-1}\frac{n_0q^2}{m}|a_{0k}|^2{\bm
k}^2,\label{DSf}\end{equation} where $a_{0k}=\int_{-\infty}^\infty
dte^{i\omega t}a_{0{\bm k}}(t),$ with $a_{0{\bm k}}(t)$ given by
Eq.~(\ref{a0k}), is the four-dimensional $(\omega,{\bm k})$
Fourier transform of the time component of the  singular part of
the  gauge field potential $a_\mu$. The explicit expression will
be obtained in the case of relatively slow dynamics of the vortex
contour. In Sec.~\ref{background}, the derivative expansion based
on the smooth contour shape is used to obtain the explicit
expression for the background part of the effective action. In the
present section, the analogous derivative expansion based on the
slow motion of the contour is the key point. Let us expose it in
detail taking the typical integral:
\begin{eqnarray}
\int_{-\infty}^\infty
\frac{d\omega}{2\pi}\cdot\frac{|f_k|^2}{\omega^2-\Omega^2_{\bm
k}+i0}&=&\int_{-\infty}^\infty
dtd\tau\frac{d\omega}{2\pi}\frac{f_{\bm k}(t)f^\ast_{\bm
k}(t+\tau)e^{i\omega\tau}}{\omega^2-\Omega^2_{\bm k}+i0}=\int
dtf_{\bm
k}(t)\sum_{l=0}^\infty\frac{(-i)^l}{l!}\frac{d^lf^\ast_{\bm
k}}{dt^l}\left[\frac{\partial^l}{\partial\omega^l}\frac{1}{\omega^2-\Omega^2_{\bm
k}}\right]_{\omega=0}=\nonumber\\&&\int dt\left[-\frac{|f_{\bm
k}(t)|^2}{\Omega^2_{\bm k}}-\frac{1}{2\Omega^4_{\bm
k}}\left(f_{\bm k}\frac{\partial^2f^\ast_{\bm k}}{\partial
t^2}+{\rm c.c.}\right)+\cdots\right],\label{tderiv}
\end{eqnarray}
where dots stand for the terms with the  higher time derivatives.
In the case of our interest, a typical $f_{\bm k}$ contains
$\dot{{\bm X}}$, hence $\partial^2f^\ast_{\bm k}/\partial t^2$,
and, consequently, the equations of motion would contain the
higher derivative term $\partial^3{\bm X}/\partial t^3$. In order
to have the equations of motion with the time derivative not
higher that two one should keep only  the first term in the
expansion Eq.~(\ref{tderiv}). Hence, in the lowest order in the
number of the derivatives over time, one finds from
Eq.~(\ref{DSf}) the correction to the effective action due to the
fluctuations:
\begin{eqnarray}
\Delta S_{\rm f}&=&\frac{\Phi^2_0}{8\pi\lambda^2_L}\int
dtd\sigma_1d\sigma_2\frac{d^3k}{(2\pi)^3}\frac{e^{i{\bm
k}\cdot{\bm X}_{21}}}{\frac{\hbar^2{\bm k}^4}{4m^2}+c^2_s{\bm
k}^2+\frac{c^2}{\lambda^2_L}}\frac{({\bm k}[\dot{{\bm
X}}_1\times{\bm X}_1^\prime])({\bm k}[\dot{{\bm X}}_2\times{\bm
X}_2^\prime])}{{\bm k}^2}.\label{DSf1}\end{eqnarray} One can see
that due to the factor $1/{\bm k}^2$ this expression is
logarithmically  divergent at large $\sigma_2-\sigma_1$. It is the
same large distance divergence as is observed in the background
action Eq.~(\ref{Sbg1}), (\ref{DSbg}), but with the opposite sign.
Let us examine the  divergent and finite  contributions in
Eq.~(\ref{DSbg}) and (\ref{DSf1}). Denoting the their sum as
$\Delta S_{\rm eff}=\Delta S_{\rm bg}+\Delta S_{\rm f}$ and taking
into account the relation $4m^2c^2_s/\hbar^2=1/\xi^2$, let us
represent $\Delta S_{\rm eff}$ in the form suitable for
integration over momentum:
\begin{eqnarray}
\Delta S_{\rm eff}&=&\frac{\Phi^2_0}{8\pi c^2}\int dt
d\sigma_2d\sigma_2[\dot{{\bm X}}_1\times{\bm
X}^\prime_1]_i[\dot{{\bm X}}_2\times{\bm
X}^\prime_2]_j\nabla_{21i}\nabla_{21j}I(\sigma_1,\sigma_2)
,\nonumber\\I(\sigma_1,\sigma_2)&=&\left\{\frac{\partial}{\partial\lambda_L^2}
\left[\lambda^2_L\int\frac{d^3k}{(2\pi)^3}\left(\frac{1}{{\bm
k}^2}-\frac{1}{{\bm k}^2+1/\lambda^2_L}\right)e^{i{\bm k}{\bm
X}_{21}}\right]-\right.\nonumber\\&&\left.\int\frac{d^3k}{(2\pi)^3}e^{i{\bm
k}{\bm X}_{21}}\left[\frac{1}{{\bm k}^2}-\frac{{\bm
k}^2+1/\xi^2}{{\bm k}^4+{\bm
k}^2/\xi^2+1/\xi^2\lambda^2_s}\right]\right\},\label{DSeff}
\end{eqnarray}
where $\nabla_{21i}=\partial/\partial X_{21i}$. The above equation
displays clearly the cancellation of the  divergent contributions
arising at $|{\bm k}|\to0$. Notice the  appearance of additional
length scale \cite{Hatsuda96}
\begin{equation}
\lambda_s=\lambda_L\cdot\frac{c_s}{c}\ll\lambda_L,\label{lambdas}\end{equation}
see Eq.~(\ref{lambdaL}), (\ref{cs}), and (\ref{xi}). This scale is
absent in the static case, but arises in the dynamical situation.
After convincing oneself that the  divergences cancel one can
write the resulting expression for the quantity
$I(\sigma_1,\sigma_2)$  in Eq.~(\ref{DSeff}):
\begin{eqnarray}
I(\sigma_1,\sigma_2)&=&-\frac{e^{-|{\bm
X}_{21}|/\lambda_L}}{4\pi|{\bm X}_{21}|}\left(1+\frac{|{\bm
X}_{21}|}{2\lambda_L}\right)+\frac{1}{8\pi|{\bm
X}_{21}|}\times\nonumber\\&&\left[e^{-|{\bm
X}_{21}|\left(1-\sqrt{1-4\xi^2/\lambda^2_s}\right)^{1/2}/\xi\sqrt{2}}
\left(1+\frac{1}{\sqrt{1-4\xi^2/\lambda^2_s}}\right)+\right.\nonumber\\&&\left.e^{-|{\bm
X}_{21}|\left(1+\sqrt{1-4\xi^2/\lambda^2_s}\right)^{1/2}/\xi\sqrt{2}}\left(1-\frac{1}{\sqrt{1-4\xi^2/\lambda^2_s}}\right)\right]
\label{int1}\end{eqnarray} at $\xi\leq\lambda_s/2$, and
\begin{eqnarray}
I(\sigma_1,\sigma_2)&=&-\frac{e^{-|{\bm
X}_{21}|/\lambda_L}}{4\pi|{\bm X}_{21}|}\left(1+\frac{|{\bm
X}_{21}|}{2\lambda_L}\right)\frac{1}{4\pi|{\bm X}_{21}|}e^{-|{\bm
X}_{21}|\sqrt{1/2\xi\lambda_s+1/4\xi^2}}\times\nonumber\\&&\left[\cos\left(|{\bm
X}_{21}|\sqrt{1/2\xi\lambda_s+1/4\xi^2}\right)+\frac{1}{\sqrt{4\xi^2/\lambda^2_s-1}}
\times\right.\nonumber\\&&\left.\sin\left(|{\bm
X}_{21}|\sqrt{1/2\xi\lambda_s+1/4\xi^2}\right)\right]
\label{int2}\end{eqnarray} at $\xi>\lambda_s/2$. Both above
expressions are interrelated by the analytical continuation. The
scale $\lambda_s$ divides the London part of the parameter space
$\xi\ll\lambda_L$ into two pieces, $\lambda_s<\xi\ll\lambda_L$ and
$\xi<\lambda_s\ll\lambda_L$. Let us consider the latter one and,
in addition, choose  a special case
$\xi\ll\lambda_s/2\ll\lambda_L$, leaving the case
$\xi>\lambda_s/2$ for a future work. Expanding square root in
Eq.~(\ref{int1}), neglecting the terms exponentially small at
$|{\bm X}_{21}|>\xi$, and applying to the resulting expression
$\nabla_{21i}\nabla_{21j}$ one finds
\begin{eqnarray}
\Delta S_{\rm eff}&=&\frac{\Phi^2_0}{32\pi^2c^2}\int dt
d\sigma_1d\sigma_2[\dot{{\bm X}}_1\times{\bm
X}^\prime_1]_i[\dot{{\bm X}}_2\times{\bm
X}^\prime_2]_j\left\{-\frac{\delta_{ij}}{|{\bm
X}_{21}|^3}\left[\left(1+\frac{|{\bm
X}_{21}|}{\lambda_s}\right)\times\right.\right.\nonumber\\&&\left.\left.e^{-|{\bm
X}_{21}|/\lambda_s}-\left(1+\frac{|{\bm
X}_{21}|}{\lambda_L}\right)e^{-|{\bm
X}_{21}|/\lambda_L}\right]+\frac{\delta_{ij}}{2\lambda^2_L|{\bm
X}_{21}|}e^{-|{\bm
X}_{21}|/\lambda_L}+\right.\nonumber\\&&\left.\frac{X_{21i}X_{21j}}{|{\bm
X}_{21}|^5}\left[\left(1+3\frac{|{\bm
X}_{21}|}{\lambda_s}+\frac{{\bm
X}_{21}^2}{\lambda^2_s}\right)e^{-|{\bm
X}_{21}|/\lambda_s}-\right.\right.\nonumber\\&&\left.\left.\left(1+3\frac{|{\bm
X}_{21}|}{\lambda_L}+\frac{{\bm
X}_{21}^2}{\lambda^2_L}\right)e^{-|{\bm
X}_{21}|/\lambda_L}+\frac{{\bm
X}_{21}^2}{2\lambda^2_L}\left(1+\frac{|{\bm
X}_{21}|}{\lambda_L}\right)e^{-|{\bm
X}_{21}|/\lambda_L}\right]\right\}\label{DSeff2}
\end{eqnarray}
Note that $\delta_{ij}\delta^3({\bm X}_{21})$ arising due to
$\nabla_{21i}\nabla_{21j}\left(1/|{\bm X}_{21}|\right)$ is
multiplied by the factor vanishing at ${\bm X}_{21}=0$ and hence
does not contribute to the effective action. The derivative
expansion adopted in the present paper, results in setting
$\sigma_2=\sigma_1+z$ and expanding all quantities depending on
$\sigma_2$ in series over $z$, see Eq.~(\ref{expansion}). In view
of the exponential damping of the integrands at large $|{\bm
X}_{21}|$ one can replace in Eq.~(\ref{DSeff2}) approximately
$\int d\sigma_1\int d\sigma_2=\int d\sigma_1\int_{-\infty}^\infty
dz$. The contributions proportional to $\delta_{ij}$ possess the
logarithmic divergence at short distances. This is an artifact due
to  the ignorance of the vortex core $|{\bm X}_{21}|\leq\xi$. One
can handle this divergence in the same manner as in
Eq.~(\ref{ultraviol}).  Taking into account another regularized
expression
\begin{equation}
\int_\xi^\infty\frac{dz}{z^3}\left(1+\frac{z}{\lambda}\right)e^{-z/\lambda}=
\frac{1}{2}\left(\frac{1}{\xi^2}+\frac{C-\ln\lambda/\xi}{\lambda^2}\right)\label{ultraviol1}
\end{equation}
where $C=0.577215....$ is the Euler constant, one can see that the
terms $\propto\ln(\lambda_L/\xi)$ cancel, the term
$\propto\ln(\lambda_s/\xi)$ survives. Gathering all the integral
together and replacing $\sigma_1\to\sigma$ one obtains the
correction to the effective action due to the combined
contributions of the excitations integrated out, and of the part
of the background action Eq.~(\ref{DSbg}):
\begin{eqnarray}
\Delta S_{\rm eff}&=&\frac{\Phi^2_0}{32\pi^2c^2}\int dt
d\sigma\left\{\left[\frac{1}{\lambda^2_s}\left(\ln\frac{\lambda_s}{\xi}-C\right)+
\kappa^2\left(\ln\frac{\lambda_L}{\lambda_s}-\frac{1}{24}\right)\right][\dot{{\bm
X}}\times{\bm X}^\prime]^2-\right.\nonumber\\&&\left.
\left(\ln\frac{\lambda_L}{\lambda_s}+\frac{1}{2}\right)
\left(\frac{\partial}{\partial\sigma}[\dot{{\bm X}}\times{\bm
X}^\prime]\right)^2+\frac{1}{2}\left(\ln\frac{\lambda_L}{\lambda_s}+\frac{3}{2}\right)(\dot{{\bm
X}}[{\bm X}^\prime\times{\bm
X}^{\prime\prime}])^2\right\}.\label{DSeff3}\end{eqnarray} The
higher derivative terms do not have the logarithmic enhancement
factors $\ln\lambda_L/\xi$ nor $\ln\lambda_s/\xi$. Hence, we
neglect them to obtain
\begin{eqnarray}
S_{\rm
eff}&\approx&\left(\frac{\Phi_0}{4\pi\lambda_L}\right)^2\int
dtd\sigma\left\{\frac{1}{2c^2_s}[\dot{{\bm X}}\times{\bm
X}^\prime]^2\ln\frac{\lambda_s}{\xi}-{\bm
X}^{\prime2}\ln\frac{\lambda_L}{\xi}\right\}+n_0q\int
d^4xa_0.\label{Sefflead}
\end{eqnarray}
The  contribution of the short range part of the electric field
has the same form $[\dot{{\bm X}}\times{\bm X}^\prime]^2$, see
Eq.~(\ref{Sbg0}). But it is not enhanced as logarithm and is
suppressed by the factor $c^2_s/c^2$. Hence, it can be  neglected.
Since $[\dot{{\bm X}}\times{\bm X}^\prime]^2=\dot{{\bm X}}^2$ in
the gauge $\dot{{\bm X}}\cdot{\bm X}^\prime=0$, ${\bm
X}^{\prime2}=1$, one can see that the first term in the braces of
Eq.~(\ref{Sefflead}) can be interpreted as the kinetic energy of
the vortex motion with the effective mass per length $L$
\begin{eqnarray}
\frac{m_{\rm
eff}}{L}&=&\left(\frac{\Phi_0}{4\pi\lambda_Lc_s}\right)^2\ln\frac{\lambda_s}{\xi}
=\frac{\pi\hbar^2}{g}\ln\frac{\lambda_s}{\xi}=
m_en_e\xi^2\times4\pi\ln\frac{\lambda_s}{\xi}.\label{meff}\end{eqnarray}
The first equality in Eq.~(\ref{meff})  coincides with the
expression obtained earlier for the straight vortex
\cite{Hatsuda96}. The last equality shows that the effective mass
of the gauge vortex equals to the mass of superconducting
electrons expelled from the region with the transverse dimension
$\xi$, multiplied by  the  dynamical enhancement factor.

\section{The local vortex equations of motion.}
\label{localeqmo} ~

First of all, let us show  that the variation of the term
$n_0q\int d^4xa_0$ in Eq.~(\ref{S0bg1}) gives nonzero result
despite the fact that the contribution of the above term to the
effective action vanishes. Using the Fourier amplitude of the
singular part of the time component of the vector potential $a_0$
[see Eq.~(\ref{a0k})] one obtains
\begin{eqnarray}
\delta\int d^4xa_0&=&-\frac{\Phi_0}{c}\int dtd\sigma(\delta{\bm
X}[\dot{{\bm X}}\times{\bm X}^\prime]).
\label{dela0}\end{eqnarray} Here, a number of integrations by
parts over both $t$ and $\sigma$ should be performed, and  the
known vector relation $[{\bm A}\times[{\bm B}\times{\bm C}]]={\bm
B}({\bm A}{\bm C})-{\bm C}({\bm A}{\bm B})$ should be taken into
account. The equations of motion of the     local (gauge) vortex
in the approximation of large logarithms can be obtained upon
varying Eq.~(\ref{S0bg1}) and (\ref{DSeff2}) over the contour
variable ${\bm X}$ and with the neglect of  the higher derivative
terms. To do so, one  should first vary over ${\bm X}$ including
the factor $e^{-|{\bm X}_{21}|/\lambda_L}$ then, second, use the
expansions Eq.~(\ref{expansion}) keeping the terms with the lowest
non-vanishing order in the expansion variable
$z=\sigma_2-\sigma_1$:
\begin{eqnarray}
\delta\int d\sigma_1d\sigma_2({\bm X}^\prime_1{\bm
X}^\prime_2)\frac{e^{-|{\bm X}_{21}|/\lambda_L}}{|{\bm
X}_{21}|}&=&2\int d\sigma_1d\sigma_2(\delta{\bm X}_1\cdot[{\bm
X}^\prime_1\times[{\bm X}_{21}\times{\bm
X}^\prime_2]])\frac{1+|{\bm X}_{21}|/\lambda_L}{|{\bm
X}_{21}|^3}e^{-|{\bm X}_{21}|/\lambda_L}=\nonumber\\&&2\int
d\sigma\int_{-\infty}^\infty
dz\frac{e^{-|z|/\lambda_L}}{2|z|}\delta{\bm X}\cdot[{\bm
X}^\prime\times[{\bm X}^\prime\times{\bm
X}^{\prime\prime}]]=-2\ln\frac{\lambda_L}{\xi}\int
d\sigma\delta{\bm X}\cdot{\bm X}^{\prime\prime}.
\end{eqnarray}
Eq.~(\ref{expansion}) should be kept in mind when deriving the
above chain of calculations.  As far as $\Delta S_{\rm eff}$ are
concerned, one can see that the terms in Eq.~(\ref{DSeff2})
arising upon contracting $[\dot{{\bm X}}_1\times{\bm
X}^\prime_1]_i[\dot{{\bm X}}_2\times{\bm X}^\prime_2]_j$ with
$X_{21i}X_{21j}$ are just the higher derivative terms which are
not  enhanced as $\ln\lambda_s/\xi$. They are displayed  in
Eq.~(\ref{DSeff3}). Using Eq.~(\ref{ultraviol}) and
(\ref{ultraviol1}) for the short distance regularization and
performing the manipulations analogous to the above, one can
obtain the variations of the terms containing velocity $\dot{{\bm
X}}$. Collecting all variations together and keeping only the
terms corresponding to the lowest derivative ones in
Eq.~(\ref{DSeff3}) one gets
\begin{eqnarray}
\delta S_{\rm
eff}&=&-\left(\frac{\Phi_0}{4\pi\lambda_L}\right)^2\int
dtd\sigma\delta{\bm X}\cdot\left\{\frac{1}{c^2_s}[{\bm
X}^\prime\frac{\partial}{\partial t}[\dot{{\bm X}}\times{\bm
X}^\prime]]\ln\frac{\lambda_s}{\xi}-{\bm
X}^{\prime\prime}\ln\frac{\lambda_L}{\xi}\right\}-\frac{n_0q\Phi_0}{c}\int
dtd\sigma(\delta{\bm X}[\dot{{\bm X}}\times{\bm X}^\prime]),
\label{deltaSeff}\end{eqnarray} which results in the equations of
motion
\begin{eqnarray}
[\dot{{\bm X}\times}{\bm X}^\prime]&=&\frac{\hbar}{2m}\left({\bm
X}^{\prime\prime}\ln\frac{\lambda_L}{\xi}+\frac{1}{c^2_s}\frac{\partial}{\partial
t}[[\dot{{\bm X}\times}{\bm X}^\prime]\times{\bm
X}^\prime]\ln\frac{\lambda_s}{\xi}\right).\label{loceqmo}
\end{eqnarray}
Let us denote for the sake of brevity
\begin{equation}\gamma=\frac{\hbar}{2m}\ln\frac{\lambda_L}{\xi}\label{gamma}\end{equation} and
consider the large amplitude motions, when the nonlinearity in
Eq.~(\ref{loceqmo}) could be essential. The zeroth order
approximate solution obtained upon neglecting the terms $\propto
1/c^2_s$ is
\begin{equation}
[\dot{{\bm X}\times}{\bm X}^\prime]^{(0)}=\gamma\kappa{\bm
n}\mbox{, }\dot{{\bm X}}^{(0)}=\gamma\kappa{\bm b}+V_l^{(0)}{\bm
X}^\prime,\label{vzero}\end{equation} where the first line in
Eq.~(\ref{frenet}) and the relation ${\bm X}^\prime=[{\bm
n}\times{\bm b}]$ are used. The longitudinal component of the
velocity $V_l^{(0)}$ cannot be determined from the first relation
in Eq.~(\ref{vzero}). However, $V_l^{(0)}$ is locally
unobservable, because of physical homogeneity of the vortex in the
(local) longitudinal direction. It can be excluded by additional
gauge choice $\dot{{\bm X}}{\bm X}^\prime=0$. So, the zeroth order
intrinsic velocity of the curved gauge vortex looks as
\begin{equation}
\dot{{\bm X}}^{(0)}={\bm b}\frac{\hbar}{2m}|{\bm
X}^{\prime\prime}|\ln\frac{\lambda_L}{\xi}.\label{v0}\end{equation}
It is proportional to the curvature of the contour and is directed
along the bi-normal vector. Its magnitude relative to the sound
velocity $c_s$ is
$$\frac{|\dot{{\bm X}}^{(0)}|}{c_s}=\frac{\hbar}{2mc_s}|{\bm
X}^{\prime\prime}|\ln\frac{\lambda_L}{\xi}=\frac{\xi}{R}\ln\frac{\lambda_L}{\xi}\ll1,$$
where $R=|{\bm X}^{\prime\prime}|^{-1}$ stands for the radius of
curvature of the contour. The large scale nonlinear motion of the
contour is rather slow. One can see that the straight vortex does
not possess an intrinsic motion as it should.

To find the correction $O(1/c^2_s)$ to the zeroth order solution
it is necessary to obtain the time derivatives of the curvature
$\kappa$ and torsion $\tau$ as well as the triple of basic vectors
of normal ${\bm n}$, bi-normal ${\bm b}$, and the tangent ${\bm
X}^\prime)$. Using Eq.~(\ref{frenet}), (\ref{vzero}),  the
definitions $\kappa=({\bm n}{\bm X}^{\prime\prime})$, $\tau=-({\bm
b}^\prime{\bm n})$ one obtains the equations
\begin{eqnarray}
\frac{\partial{\bm X}^\prime}{\partial
t}&=&\gamma(\kappa^\prime{\bm
b}-\kappa\tau{\bm n}),\nonumber\\
\frac{\partial{\bm n}}{\partial t}&=&\gamma\left[\kappa\tau{\bm
X}^\prime+\left(\frac{\kappa^{\prime\prime}}{\kappa}-\tau^2\right){\bm
b}\right],\nonumber\\
\frac{\partial{\bm b}}{\partial
t}&=&-\gamma\left[\kappa^\prime{\bm
X}^\prime+\left(\frac{\kappa^{\prime\prime}}{\kappa}-\tau^2\right){\bm
n}\right]\label{vectimederiv}
\end{eqnarray}
and
\begin{eqnarray}
\frac{\partial\kappa}{\partial
t}&=&-\gamma(2\kappa^\prime\tau+\kappa\tau^\prime),\nonumber\\
\frac{\partial\tau}{\partial
t}&=&\gamma\left[\kappa\kappa^\prime+\left(\frac{\kappa^{\prime\prime}}{\kappa}-\tau^2\right)^\prime\right].
\label{kttimederiv}\end{eqnarray} Equations (\ref{kttimederiv})
(however, without the factor $\gamma$) are known in hydrodynamics
as Da Rios equations \cite{Ricca}. Making use of
Eq.~(\ref{frenet}), (\ref{loceqmo}), (\ref{vzero}),
(\ref{vectimederiv}), and (\ref{kttimederiv}) one can obtain the
relation for finding  the correction $\dot{{\bm X}}^{(1)}$ due to
the term $O(1/c^2_s)$:
\begin{eqnarray}
[\dot{{\bm X}}^{(1)}\times{\bm
X}^\prime]&=&\frac{\hbar\gamma^2}{2mc^2_s}\ln\frac{\lambda_s}{\xi}
\left[{\bm n}(\kappa^{\prime\prime}-\kappa\tau^2)+{\bm
b}(2\kappa^\prime\tau+\kappa\tau^\prime)\right].
\end{eqnarray}
Using the above relation one gets
\begin{eqnarray}
\dot{{\bm
X}}^{(1)}&=&\frac{\hbar\gamma^2}{2mc^2_s}\ln\frac{\lambda_s}{\xi}
\left[{\bm b}(\kappa^{\prime\prime}-\kappa\tau^2)-{\bm
n}(2\kappa^\prime\tau+\kappa\tau^\prime)\right]=|\dot{{\bm
X}}^{(0)}|\xi^2\ln\frac{\lambda_L}{\xi}\ln\frac{\lambda_s}{\xi}\times\nonumber\\&&
\left[{\bm b}(\kappa^{\prime\prime}-\kappa\tau^2)-{\bm
n}(2\kappa^\prime\tau+\kappa\tau^\prime)\right]\frac{1}{\kappa}.
\label{v1}\end{eqnarray} The ratio of correction to the zeroth
order velocity
$$\frac{|\dot{{\bm X}}^{(1)}|}{|\dot{{\bm
X}}^{(0)}|}=
\xi^2\kappa^2\ln\frac{\lambda_s}{\xi}\ln\frac{\lambda_L}{\xi}<1$$
is small but non-negligible, because the truly small  ratio of the
coherence length to the curvature radius squared is enhanced by
the product of two large logarithms.

Let us consider the small amplitude oscillations around the
locally straight vortex. Then, locally, one may choose ${\bm
X}^\prime={\bm e}_z$, ${\bm X}(\sigma,t)={\bm u}(\sigma,t)+{\bm
e}_z\sigma$, where ${\bm u}$ is orthogonal to the contour
direction ${\bm e}_z$: ${\bm e}_z\cdot{\bm u}(\sigma,t)=0$.
Introducing the characteristic velocity
\begin{equation}
c^2_0=c^2_s\times\frac{\ln(\lambda_L/\xi)}{\ln(\lambda_s/\xi)},\label{c0}\end{equation}
and restricting oneself in Eq.~(\ref{loceqmo}) by the linear
approximation, one obtains the equation
\begin{equation}
\frac{\ddot{{\bm u}}}{c^2_0}-{\bm u}^{\prime\prime}
+\frac{1}{\gamma}[\dot{{\bm u}}\times{\bm
e}_z]=0\label{loceqmolin}\end{equation}for the two-dimensional
vector ${\bm u}=(u_x,u_y)$. The plain wave solution
$u_{x,y}\propto e^{i(k\sigma-\omega t)}$ corresponds to the
oscillation spectrum
$$\omega=\frac{c^2_0}{2\gamma}\left(\pm1\pm\sqrt{1+\frac{4\gamma^2k^2}{c^2_0}}\right),$$and
the relations among the amplitudes $u_y=\pm iu_x$. In our case,
$\gamma^2k^2/c^2_0=2k^2\xi^2\ln(\lambda_L/\xi)\ln(\lambda_s/\xi)\ll1$,
because the typical wave numbers $k\ll1/\xi$, so that the square
root can be expanded. Then the general solution of
Eq.~(\ref{loceqmolin}), assuming as usual the periodic boundary
conditions, can be represented in the form of the superposition of
the left and right circularly polarized waves:
\begin{equation}
{\bm u}(\sigma,t)=\frac{{\bm e}_x+i{\bm
e}_y}{\sqrt{2}}\sum_{l=-\infty}^\infty\left(\alpha_le^{-i\omega_1t}+\beta_le^{-i\omega_2t}\right)
e^{ik_l\sigma}+{\rm c.c}.\end{equation} Here, $k_l=2\pi l/L$, $L$
stands for the periodicity length,  $l=0,1,2...$,
\begin{eqnarray}
\omega_1&=&\gamma k^2_l=\frac{\hbar
k^2_l}{2m}\ln\frac{\lambda_L}{\xi},\nonumber\\
\omega_2&=&\frac{c^2_0}{\gamma}+\gamma
k^2_l=\frac{2n_0g}{\hbar\ln(\lambda_s/\xi)}+\frac{\hbar
k^2_l}{2m}\ln\frac{\lambda_L}{\xi},\label{omegas}\end{eqnarray}
and $\alpha_l$, $\beta_l$ being the arbitrary complex constants.
The first branch $\omega_1$ coincides with the Friedel-De
Gennes-Matricon waves \cite{friedel,fetter2}, if one allows for
the fact that here the scalar field mimics the condensate of the
Cooper pairs with the mass $m=2m_e$ and charge $q=2e$. It
originates from the fact that the vortex segment being accelerated
by the tension force $\propto{\bm X}^{\prime\prime}$ acquires
velocity $\Delta{\bm v}\propto{\bm n}$ in the normal vector
direction. The Magnus-like force $\propto[{\bm e}_z\times\dot{{\bm
u}}]$ results in the circular motion of the segment in the plane
perpendicular to ${\bm e}_z$ and ${\bm n}$, that is in the
direction of bi-normal vector ${\bm b}=[{\bm X}^\prime\times{\bm
n}]$. The high frequency branch $\omega_2$ possesses the gap
$$\omega_{\rm min}=\frac{2n_0g}{\hbar\ln(\lambda_s/\xi)}.$$
It can be interpreted as follows. Since $\omega_{\rm
min}=\omega_2(k_l=0)$, this mode corresponds to the neglect of the
term $\propto{\bm X}^{\prime\prime}$ in the vortex equations of
motion Eq.~(\ref{loceqmolin}). Using Eq.~(\ref{meff}) for the
effective vortex mass together with Eq.~(\ref{cs}), (\ref{gamma}),
and (\ref{c0}) one can rewrite the equations of motion in this
limiting case as
\begin{equation}
\frac{m_{\rm eff}}{L}\ddot{{\bm u}}=2\pi\hbar n_0[{\bm
e}_z\times\dot{{\bm u}}],\label{magnus}\end{equation} which is the
dissipation- and pinning-free part of the vortex equation of
motion including the transverse force \cite{ao03}.

The characteristic velocity $c_0$ in the case of gauge vortex
exceeds the sound velocity $c_s$ [see Eq.~(\ref{c0})], hence both
branches of small transverse oscillations are expected to be
damped due to the emission of the Bogolyubov-Anderson excitations
(sound waves in the large wave length limit). The calculation of
the damping rate goes beyond the scope of the present paper.

\section{The case of the global vortex.}
\label{global} ~

The global  vortex can be treated in the same manner. The
corresponding field theoretic model can be obtained from AHM one
upon switching off the gauge field degrees of freedom and  is
called the Goldstone model \cite{goldstone} in particle physics
and the Gross-Pitaevskii one \cite{Pit,Gro} in condensed matter
physics. Omitting the chain of derivations completely analogous to
those presented in proceeding sections let us write the effective
action $S_{\rm eff}=S_{\rm bg}+\Delta S_{\rm f}$ of the global
model as the sum of the action of the background field $S_{\rm
bg}$ and that of the integrated out fluctuations of the modulus
and the phase of the scalar field $\Delta S_{\rm f}$. Here $S_{\rm
bg}$ looks like
\begin{equation}
S_{\rm bg}=\hbar n_0\int d^4x\left[a_0-\partial_t\chi_{\rm
b}-\frac{\hbar}{2m}({\bm a}+{\bm\nabla}\chi_{\rm
b})^2\right],\label{Sbgg}
\end{equation}
and $\chi_{\rm b}$ stands for the smooth background phase
representing the possible potential flow whose velocity is ${\bm
v}_{\rm b}=\frac{\hbar}{m}{\bm\nabla}\chi_{\rm b}$, while
$a_0=-\partial_t\chi_s$ and ${\bm a}={\bm\nabla}\chi_s$ represent
the time and space derivatives of the singular phase $\chi_s$
responsible for the vortex. Their space Fourier components are
given by the equations analogous to Eq.~(\ref{a0k}) and
(\ref{ak}), respectively:
\begin{eqnarray}
a_{0{\bm k}}(t)&=&-\frac{2\pi i}{{\bm k}^2}\sum_an_a\int
d\sigma_a{\bm k}[\dot{{\bm X}}_a\times{\bm X}^\prime_a]e^{-i{\bm
k}{\bm X}_a},\nonumber\\
{\bm a}_{{\bm k}}(t)&=&\frac{2\pi i}{{\bm k}^2}\sum_an_a\int
d\sigma_a[{\bm k}\times{\bm X}^\prime_a]e^{-i{\bm k}{\bm X}_a}.
\label{amukg}
\end{eqnarray}
From now on let us restrict ourselves by the single  vortex with
the unit quantum of circulation $n_a=1$. Then the coordinate space
expressions  found from Eq.~(\ref{amukg}) by the inverse Fourier
transform are
\begin{eqnarray}
a_0(t,{\bm x})&=&\frac{1}{2}\int d\sigma\frac{({\bm x}-{\bm
X})\cdot[\dot{{\bm X}}\times{\bm X}^\prime]}{|{\bm x}-{\bm
X}|^3},\nonumber\\
{\bm a}(t,{\bm x})&=&\frac{1}{2}\int d\sigma\frac{[{\bm
X}^\prime\times({\bm x}-{\bm X})]}{|{\bm x}-{\bm X}|^3},
\label{amurg}
\end{eqnarray}
${\bm X}\equiv{\bm X}(t,\sigma)$. The term originating from the
integrated out fluctuations can be obtained from Eq.~(\ref{DSf})
by taking the limit of vanishing   gauge coupling constant $q$ and
restoring the term $\propto{\bm a}^2/2m$ omitted when discussing
the  gauge vortex:
\begin{equation}
\Delta S_{\rm f}=\frac{\hbar^2n_0}{2m}\int\frac{d^3k}{(2\pi)^3}
\frac{|\rho_{\bm k}|^2}{\hbar^2{\bm k}^2/4m^2+c^2_s},\label{DSfg}
\end{equation}
where, for the sake of brevity, we introduce the quantity
$\rho_{\bm k}=\int d^3x\rho e^{-i{\bm k}\cdot{\bm r}}$, which is
the Fourier amplitude of the quantity
\begin{equation}
\rho=a_0-\partial_t\chi_{\rm b}-\frac{\hbar}{2m}({\bm
a}+{\bm\nabla}\chi_{\rm b})^2. \label{rho}
\end{equation}
Notice that this $\rho$ (taken at $a_0=0$, ${\bm a}=0$, that is,
in the case of no vortex present) is just that combination of the
derivatives of the Goldstone field $\chi_{\rm b}$ which was
introduced in Ref.~\cite{schakel,gww} on the basis of the Galilean
invariance of the effective action. Since the typical momenta are
subjected to the condition $|{\bm k}|\ll1/\xi$, where the healing
length $\xi$ is given by Eq.~(\ref{xi}), the correction
Eq.~(\ref{DSfg}) to the background action can be represented as
the series in the number of derivatives
\begin{equation}
\Delta S_{\rm f}=\frac{\hbar^2n_0}{2mc^2_s}\int
dtd^3x\left[\rho^2-\xi^2\sum_{l=0}^\infty(\xi^2{\bm\nabla}^2)^l({\bm\nabla}\rho)^2\right].
\label{DSfg1}
\end{equation}
The leading order effective action in the case of no vortex
present is, omitting the total derivative term,
\begin{equation}
S_{\rm eff}=\frac{\hbar^2n_0}{2m}\int
d^4x\left[\frac{1}{c^2_s}(\partial_t\chi_{\rm
b})^2-({\bm\nabla}\chi_{\rm b})^2+...\right],
\end{equation}
which is just the action of the Goldstone mode in neutral
superfluid. This agrees with the general analysis \cite{gww},
where the coefficient in the effective action was fixed on the
basis of the equations of motion of the Goldstone field, about
which it is known that this field is the sound wave.

Turning back to deriving the global vortex equations of motion let
us set the smooth phase $\chi_{\rm b}$ to zero and neglect in
Eq.~(\ref{DSfg1}) all terms $\propto\xi^2$. When finding the
variation of the background effective action Eq.~(\ref{Sbgg}) over
${\bm X}$ the following expressions derived from Eq.~(\ref{amukg})
are necessary:
\begin{eqnarray}
\delta a_0({\bm x})&=&-\int\frac{d^3k}{(2\pi)^2}e^{i{\bm
k}\cdot{\bm x}}\int d\sigma e^{-i{\bm k}\cdot{\bm
X}}\left\{\delta{\bm X}\cdot[\dot{{\bm X}}\times{\bm
X}^\prime]+\frac{\delta{\bm X}\cdot[{\bm k}\times{\bm
X}^\prime]}{{\bm
k}^2}i\partial_t\right\},\nonumber\\
\delta{\bm a}({\bm x})&=&\int\frac{d^3k}{(2\pi)^2}e^{i{\bm
k}\cdot{\bm x}}\int d\sigma e^{-i{\bm k}\cdot{\bm
X}}\left\{[\delta{\bm X}\times{\bm X}^\prime]-\frac{{\bm k}({\bm
k}\cdot[\delta{\bm X}\times{\bm X}^\prime])}{{\bm
k}^2}\right\}.\label{damukg}
\end{eqnarray}
In fact, the time derivative term in $\delta a_0$ in
Eq.~(\ref{damukg}) vanishes when applied  to the background
action. The zeroth order equations of motion resulting from the
variation of the background action are
\begin{equation}
[\dot{{\bm X}}\times{\bm X}^\prime]-\frac{\hbar}{m}[{\bm a}({\bm
X})\times{\bm X}^\prime]=0.\label{globeqmo}
\end{equation}
One can see from Eq.~(\ref{globeqmo}) that upon neglecting the
integrated out fluctuations of the phase and the modulus of the
scalar field  the classic law of the vortex motion
\begin{equation}
\dot{{\bm X}}=\frac{\hbar}{m}{\bm a}({\bm X})={\bm
V}\label{hydro}\end{equation} is recovered,  because
$\frac{\hbar}{m}{\bm a}({\bm X})$ is the velocity induced at the
given vortex location ${\bm x}={\bm X}(\sigma)$ by other elements
of the vortex. Inserting here Eq.~(\ref{amurg}), using
Eq.~(\ref{expansion}), and taking the low and upper integration
limits to be, respectively, $\xi$ and $R$ (see below), one obtains
in the zeroth order
\begin{eqnarray}
\dot{{\bm X}}^{(0)}(\sigma_1)&=&\frac{\hbar}{m}{\bm a}({\bm
X})=\frac{\hbar}{2m}\int d\sigma_2\frac{[{\bm X}_{21}\times{\bm
X}^\prime_2]}{|{\bm X}_{21}|^3}\approx\frac{\hbar}{2m}[{\bm
X}^\prime_1\times{\bm
X}^{\prime\prime}_1]\ln\frac{R}{\xi}\equiv\gamma_g\kappa{\bm b},
\label{V0g}
\end{eqnarray}
where ${\bm b}$ is the bi-normal vector, and
\begin{equation}
\gamma_g=\frac{\hbar}{2m}\ln\frac{R}{\xi}.\label{gammag}
\end{equation}Recall that ${\bm X}_{1,2}\equiv{\bm
X}(\sigma_{1,2})$, and $\kappa=|{\bm X}^{\prime\prime}|$. The
result (\ref{V0g}) coincides with the velocity of the vortex ring
in superfluid HeII \cite{LifPit}. The large-scale motions of the
contour in the case of the global vortex is  slow, because
$$\frac{|\dot{{\bm
X}}^{(0)}|}{c_s}=\frac{\hbar\kappa}{2mc_s}\ln\frac{R}{\xi}\sim\frac{\xi}{R}\ln\frac{R}{\xi}\ll1.$$
Here the curvature of the contour $\kappa$ is approximated by the
inverse cutoff $1/R$.

As far as $\Delta S_{\rm f}$ are concerned, we evaluate its
contribution under the same assumption $|{\bm k}|\ll1/\xi$ as made
when discussing the local vortex case. However, in addition to the
short distance divergence at $r\leq\xi$ the energy per unit length
of global vortex has the pertinent divergence at large distances
due to the real gapless  Goldstone field.  As usual, one
regularizes it by means of the  integration cutoff  at the
distances $r\sim R$, where $R$ is of the order of the curvature
radius of the contour. We make here the large distance
regularization by means of the large distance  cutoff ${\bm
k}^{-2}\to({\bm k}^2+1/R^2)^{-1}$, $R\to\infty$. Performing the
integration over momentum  and making the same approximation as in
the case of the local vortex one finds
\begin{eqnarray}
\Delta S_{\rm f}&\approx&\frac{\pi\hbar^2n_0}{2mc^2_s}\int
dtd\sigma\left\{[\dot{{\bm X}}\times{\bm
X}^\prime]^2\ln\frac{R}{\xi}+\frac{3R^2}{4}(\dot{{\bm
X}}\cdot[{\bm X}^\prime\times{\bm X}^{\prime\prime}])^2\right\}.
\label{DSfg3}
\end{eqnarray}
Contrary to the case of magnetic flux vortex, here the higher
derivative term is multiplied by the large factor $R^2$. The ratio
of the second higher derivative term in the braces to the first
lowest order derivative term is estimated with the help of
Eq.~(\ref{V0g}) to be
$$\frac{R^2\kappa^2}{\ln R/\xi}\sim\frac{1}{\ln R/\xi}\ll1,
$$where we take the cutoff $R\sim1/\kappa$. So, the higher
derivative here in the case of the global vortex is also
suppressed, although without additional small factor
$\lambda^2_L\kappa^2$ pertinent to the gauge vortex case. As in
Sec.~\ref{localeqmo}, when obtaining the correction to the
equations of motions, the higher derivative term is neglected.
Analogously to the gauge vortex case, here the lowest derivative
term in the action can be interpreted as the contribution of the
kinetic energy of the vortex segment with the effective mass per
length
\begin{equation}
\frac{m_{\rm
eff}}{L}=\frac{\pi\hbar^2n_0}{mc^2_s}\ln\frac{R}{\xi}=\frac{\pi\hbar^2}{g}\ln\frac{R}{\xi}.
\label{meffg}
\end{equation}
This coincides with the expressions obtained in
Ref.~\cite{wex96,Hatsuda96}, allowing for apparent typo (extra
mass in the denominator) in Ref.~\cite{wex96}.

The  correction to the zeroth approximation should be derived upon
varying the nonlocal form of the action Eq.~(\ref{DSfg3}). Taking
into account the zeroth order equations of motion, one finds
\begin{equation}
[\dot{{\bm X}}\times{\bm X}^\prime]=\frac{\hbar}{m}[{\bm a}({\bm
X})\times{\bm
X}^\prime]+\frac{\hbar}{2mc^2_s}\ln\frac{R}{\xi}[\partial_t[\dot{{\bm
X}}\times{\bm X}^\prime]\times{\bm X}^\prime]. \label{Vg}
\end{equation}
Recalling that ${\bm a}({\bm
X})\approx\frac{\kappa}{2}\ln\frac{R}{\xi}{\bm b}$ [see
Eq.~(\ref{V0g})] Eq.~(\ref{Vg}) looks similar to
Eq.~(\ref{loceqmo}). Hence the correction due to excitations of
the modulus and the phase of the scalar field can obtained from
Eq.~(\ref{v1}) by means of  the replacements $\gamma\to\gamma_g$,
$\lambda_s,\lambda_L\to R$:
\begin{eqnarray}
\dot{{\bm
X}}^{(1)}&=&\frac{\hbar\gamma^2_g}{2mc^2_s}\ln\frac{R}{\xi}
\left[{\bm b}(\kappa^{\prime\prime}-\kappa\tau^2)-{\bm
n}(2\kappa^\prime\tau+\kappa\tau^\prime)\right]=\nonumber\\&&|\dot{{\bm
X}}^{(0)}|\xi^2\ln^2\frac{R}{\xi}\left[{\bm
b}(\kappa^{\prime\prime}-\kappa\tau^2)-{\bm
n}(2\kappa^\prime\tau+\kappa\tau^\prime)\right]\frac{1}{\kappa}.
\label{v1g}\end{eqnarray}The ratio of the correction to the zeroth
order approximation is small,
$$\frac{|\dot{{\bm X}}^{(1)}|}{|\dot{{\bm
X}}^{(0)}|}= \xi^2\kappa^2\ln^2\frac{R}{\xi}<1,$$ but
non-negligible, because of the large logarithm squared. As in the
case of the magnetic vortex, the large scale nonlinear motion is
slow.

The linear  small amplitude motion  of the global vortex is
analyzed by taking the limit of locally straight contour, see
preceding  section for more detail. The corresponding equations of
motion looks similar to Eq.~(\ref{loceqmolin}), in which one
should make the replacements $c_0\to c_s$, because the ratio of
logarithms in Eq.~(\ref{c0}) drops in the global vortex limit, and
$\gamma\to\gamma_g$. Hence,  the small transverse motions of the
global vortex are characterized by the velocity of sound which
means that the kinematics is exactly on the threshold of  the
emission of sound waves, and the phase space of the emitted waves
shrinks to zero. By this reason the small amplitude motions are
not expected to be damped. The oscillation spectrum is given by
Eq.~(\ref{omegas}) with the proper replacements just mentioned,
together with $\lambda_L,\lambda_s\to R$. The lower branch of the
spectrum is well-known as the Kelvin waves
\cite{Pit,LifPit,fetter1}. The branch with the gap $$\omega_{\rm
min}=\frac{2n_0g}{\hbar\ln(R/\xi)}$$ is interpreted similarly to
that of the local case  as being due to the acceleration of the
vortex segment with the mass per length Eq.~(\ref{meffg})
subjected to the Magnus force. The equations of motion look the
same as Eq.~(\ref{magnus}), with  the effective mass
Eq.~(\ref{meff}) replaced by  one in Eq.~(\ref{meffg}).

\section{Discussion.}
\label{discus}~

The usual attitude to the effective lagrangians   is that they do
not refer to underlying basic theory  and its fundamental
constituents but are formulated in terms of effective degrees of
freedom. In the case of condensed mater physics they are the
modulus and phase of the order parameter together with the
screened electromagnetic field in case of charged superfluid, and
the same except the electromagnetic fields in case of neutral
superfluids.  Guided by analogous approach based on the field
theoretic models possessing the property of spontaneous symmetry
breaking we obtain here the effective actions and the equations of
motion to the lowest number of derivatives over contour parameter
and time for  the local (gauge) and global vortices. Of course,
the models used in the present paper are idealized in that they
ignore the friction and pinning effects, but the inertial
properties of the vortex state are most conveniently treated in
the framework of such models. The key feature of the present
consideration is the usage of the explicitly transverse electric
gauge field strength in the study of the gauge vortex dynamics. If
the field excitations were not included, the resulting effective
action of the background field configuration would contain
divergence at large distance. The crucial role of the excitations
is that they cancel this divergence so that the gauge vortex
dynamics becomes finite at large distances as it should, because
all gauge fields in spontaneously broken gauge model are screened
at the distances larger than the penetration depth. As a
consequence, the higher derivative terms in the effective action
for curved gauge vortex are small.  Taking into account the scalar
field excitations in the global models like the Gross-Pitaevskii
or the Goldstone one  results in the correct Galilean-invariant
form of effective action for the Goldstone mode suggested in
Ref.~\cite{gww} on the symmetry grounds.

I am grateful to G.~I.~Shvetsova for encouragement and warm
support.

\end{document}